\documentclass[12pt]{article}
\usepackage[T1]{fontenc}
\usepackage[utf8]{inputenc}
\usepackage{ulem}
\usepackage{color}
\usepackage{bm}
\usepackage{slashed}
\usepackage{varioref}
\usepackage[usenames,dvipsnames]{xcolor}
\usepackage{graphicx}
\usepackage{amsmath}
\usepackage{amssymb}
\usepackage{amsfonts}
\usepackage{amsthm}
\usepackage{array}
\usepackage{yfonts}
\usepackage{float}
\usepackage{comment}
\usepackage{caption}
\usepackage[sort]{cite}
\usepackage{subcaption}
\usepackage{wrapfig}
\usepackage[
      colorlinks=true,
      linkcolor=blue,
      urlcolor=blue,
      filecolor=blue,
      citecolor=red,
      pdfstartview=FitV,
      pdftitle={},
      pdfauthor={},
      pdfsubject={},
      pdfkeywords={},
      bookmarksopen=true
]{hyperref}

\usepackage{epsfig}

\usepackage{hyperref}
\newcommand{\mcN}{{\mycal N}}

\newcommand{\sectionofScri}%
{{ \,\,\,\,\mathring{\!\!\!\!\mcN}}}

\newcommand{\scri}{{\mycal I}}%
\newcommand{\eeal}[1]{\label{#1}\end{eqnarray}}

\DeclareFontFamily{OT1}{rsfs}{}
\DeclareFontShape{OT1}{rsfs}{m}{n}{ <-7> rsfs5 <7-10> rsfs7 <10-> rsfs10}{}
\DeclareMathAlphabet{\mycal}{OT1}{rsfs}{m}{n}

\newcommand{\mmr}[1]{\mnotex{{\bf mm:} {\color{violet} #1}}}

\definecolor{applegreen}{rgb}{0.55, 0.71, 0.0}
\definecolor{armygreen}{rgb}{0.29, 0.33, 0.13}

\definecolor{caribbeangreen}{rgb}{0.0, 0.8, 0.6}

\newcounter{mnotecount}[section]

\renewcommand{\themnotecount}{\thesection.\arabic{mnotecount}}

\newcommand{\mnotex}[1]
{\protect{\stepcounter{mnotecount}}$^{\mbox{\footnotesize
        $
        \bullet$\themnotecount}}$ \marginpar{
    \raggedright\tiny\em
    $\!\!\!\!\!\!\,\bullet$\themnotecount: #1} }

\newcommand{\ptc}[1]{\mnotex{{\bf ptc:} {  #1}}}
\newcommand{\ptcrr}[1]{\mnotex{{\bf ptc:} {  #1}}}

\newcommand{\bel}[1]{\begin{equation}\label{#1}}

\newcommand{\eeq}{\end{equation}}
\newcommand{\ee}{\end{equation}}
\newcommand{\beqa}{\begin{eqnarray}}
\newcommand{\eeqa}{\end{eqnarray}}
\newcommand{\beqan}{\begin{eqnarray*}}
\newcommand{\eeqan}{\end{eqnarray*}}
\newcommand{\ba}{\begin{array}}
\newcommand{\ea}{\end{array}}

\newcommand{\ptcheck}[1]{\ptc{checked on #1}}

\newcommand{\sey}[1]{{\color{orange} #1}}

\def\beq{\begin{eqnarray}}
\def\eeq{\end{eqnarray}}

\def\be{\begin{equation}}
\def\ee{\end{equation}}
\def\bea{\begin{eqnarray}}
\def\eea{\end{eqnarray}}

{\catcode `\@=11 \global\let\AddToReset=\@addtoreset}
\AddToReset{equation}{section}

\newcommand{\jhbr}[1]{\mnotex{{\bf jh:} {\color{blue} #1}}}

\newcommand{\tscomr}[1]{\mnotex{{\bf ts:} {  #1}}}

{\catcode `\@=11 \global\let\AddToReset=\@addtoreset}
\AddToReset{equation}{section}

\renewcommand{\ptcrr}[1]{}

\renewcommand{\mmr}[1]{}
\renewcommand{\tscomr}[1]{}

\renewcommand{\ptcheck}[1]{}
\renewcommand{\jhbr}[1]{}

\begin{document}

\title{\bf{The $SO(1,4)$ flux-balance laws \\ of de Sitter at quadrupolar order}}

\date{}

\maketitle

\vspace{-1.5cm}

\centerline{\large{\bf{Geoffrey 
Comp\`ere}$^{a}$\footnote{geoffrey.compere@ulb.be},  Sk Jahanur 
Hoque$^{a,b,c}$\footnote{jahanur.hoque@hyderabad.bits-pilani.ac.in}, Emine Seyma Kutluk$^{d,e}$\footnote{emine.kutluk@sns.it}}}\vspace{6pt}

\bigskip\medskip
\centerline{\textit{{}$^{a}$ Universit\'{e} Libre de Bruxelles, International Solvay Institutes,}}
\centerline{\textit{CP 231, B-1050 Brussels, Belgium}}

\medskip
\centerline{\textit{{}$^{b}$ Birla Institute of Technology and Science, Pilani, Hyderabad Campus, }}
\centerline{\textit{Jawaharnagar, Hyderabad 500 078, India}}

\medskip
\centerline{\textit{{}$^{c}$ Institute of Theoretical Physics, Faculty of Mathematics and Physics, Charles University, }}
\centerline{\textit{V Holešovičkách 2, 180 00 Prague 8, Czech Republic}}

\medskip
\centerline{\textit{{}$^{d}$ Scuola Normale Superiore,}}
\centerline{\textit{Piazza dei Cavalieri 7, Pisa, 56126, Italy}}

\medskip
\centerline{\textit{{}$^{e}$ INFN Pisa,}}
\centerline{\textit{Largo Pontecorvo 3, 56127 Pisa, Italy}}

\vspace{1cm}


\begin{abstract}
The linear solution for quadrupolar perturbations around de Sitter spacetime was recently constructed. 
In this paper, we provide the flux-balance laws for each background symmetry (dilatations, rotations, spatial translations and cosmological boosts) in terms of source moments at quadrupolar order. We write the dilatation flux-balance law in two distinct ways, which allows to contrast two distinct proposals for the negative definite energy flux. The standard Poincar\'e flux balance laws at future null infinity are recovered in the flat limit of the $SO(1,4)$ flux-balance laws. 
\end{abstract}
\newpage
\tableofcontents

\section{Introduction}

According to the standard model of cosmology, our universe admits a positive cosmological constant $\Lambda \approx 3 \times 10^{-122}$ in Planck units \cite{SupernovaSearchTeam:1998fmf,SupernovaCosmologyProject:1998vns}. Though tiny, the effect of the cosmological constant dominates in the infrared and leads to an asymptotically de Sitter spacetime structure at late times, which overarch in particular the definition of asymptotically conserved quantities. Besides, gravitational wave observations have provided within the last decade new insights into astrophysics and cosmology \cite{LIGOScientific:2016aoc}. These observations have led to a large theoretical effort to define gravitational waves emitted from localized sources in a cosmological background, see e.g.  \cite{Ashtekar:2014zfa,Bohmer:2014sjm,Chu:2015yua,Ashtekar:2015lxa,Ashtekar:2015ooa,Szabados:2015wqa,Date:2015kma,Bishop:2015kay,Chrusciel:2016oux,Saw:2016isu,Kesavan:2016pvc,Saw:2016pez,Schlue:2016nlb,Date:2016uzr,Viaggiu:2017xoh,Hamada:2017gdg,Hoque:2017ewa,Saw:2017zks,Saw:2017ziu,Saw:2017amv,Ashtekar:2017wgq,Ashtekar:2017ydh,Bonga:2017dpl,Ashtekar:2017dlf,Viaggiu:2017eeg,Bonga:2017dlx,Hoque:2018byx,Szabados:2018erf,Poole:2018koa,Mao:2019ahc,PremaBalakrishnan:2019jvz,Ashtekar:2019khv,Compere:2019bua,Ruzziconi:2019pzd,Chrusciel:2020rlz,Compere:2020lrt,Fernandez-Alvarez:2020hsv,Kolanowski:2020wfg,Bonga:2020fhx,Fier:2021fbt,Chrusciel:2021ttc,Kolanowski:2021hwo,Fier:2021rud,Chakraborty:2021ezq,Fiorucci:2021pha,Poole:2021avh,Dobkowski-Rylko:2022dva,Geiller:2022vto,Kaminski:2022tum,Grant:2022koa,Senovilla:2022pym,Bonga:2023eml,Fernandez-Alvarez:2023wal,Compere:2023ktn,Harsh:2024kcl,Fernandez-Alvarez:2024bkf,Dobkowski-Rylko:2024jmh}.

Many simple theoretical questions remain unanswered. In this article, we will concentrate our focus on two questions: What is the role of the $SO(1,4)$ de Sitter symmetry regarding gravitational radiation? What is the definition of energy in asymptotically de Sitter spacetime? 

It was demonstrated in \cite{Compere:2019bua,Compere:2020lrt} that even though gravitational waves change the leading components of the metric at future infinity, a boundary gauge exists such the asymptotic symmetry algebra of asymptotically de Sitter spacetime is the infinite-dimensional $\Lambda$-BMS algebra, which reduces to the BMS algebra \cite{1962RSPSA.269...21B,1962PhRv..128.2851S} in the flat limit. The structure constants of the $\Lambda$-BMS algebra generically depend upon the radiation field. However, at quadratic order in perturbation theory around de Sitter, one can isolate a universal $SO(1,4)$ algebra of background symmetries, as we will review in Section \ref{sec:charges}, simply because the fluxes are already quadratic in the fields once the symmetry generators are taken to be the background Killing vectors of de Sitter. Our first objective is to derive the set of  flux-balance laws associated with the background $SO(1,4)$ symmetry and evaluate them in the physical phase space. We will perform this computation at the level of the quadrupolar truncation of the linear spectrum \cite{Compere:2023ktn}, which has been demonstrated to match \cite{Harsh:2024kcl} with the spectrum of perturbations obtained from independent derivations \cite{Bonga:2023eml,Harsh:2024kcl}. 

Several formulae for the energy flux in asymptotically de Sitter spacetime have been proposed \cite{Kastor:2002fu,Luo:2007se,Kelly:2012zc,Szabados:2015wqa,Zhang:2015iua,Chrusciel:2016oux,Saw:2016isu,Saw:2017zks,Saw:2017ziu,Saw:2017amv,Bonga:2017dlx,Hoque:2018byx,Szabados:2018erf,PremaBalakrishnan:2019jvz,Chrusciel:2020rlz,Kolanowski:2020wfg,Dobkowski-Rylko:2022dva,Senovilla:2022pym,Dobkowski-Rylko:2024jmh}. Our second objective is to identify from the dilatation flux-balance law, how one can identify a charge and a flux such that (i) the flux reduces to the standard energy flux in the flat limit, (ii) the flux is manifestly non-negative at quadratic order. We will contrast our expressions for the energy loss and angular momentum flux with part of the literature. The linear momentum and boost charge loss formulae that we will derive have not yet appeared in that form in the literature to the best of our knowledge. 

The rest of the paper is organized as follows. We first define the $SO(1,4)$ flux-balance laws in perturbation theory around de Sitter spacetime from first principles in Section \ref{sec:charges}. We then evaluate them on the solution space of even and odd quadratic perturbations of de Sitter in Section \ref{sec:eval} and we prove that the standard flat spacetime limit is obtained. We compare our expressions of the energy loss and angular momentum loss with the literature in Section \ref{sec:comp}. We finally conclude in Section \ref{sec:ccl}.

\section{Charges and Fluxes on de Sitter space-times}
\label{sec:charges}

The future region $\scri^+$ of asymptotically de Sitter spacetimes with cosmological constant $\Lambda=3H^2$ admits a Starobinsky expansion  \cite{Starobinsky:1982mr} of the form 
\begin{equation}
ds^2 = -\frac{d\tau^2}{H^2\tau^{2}}  + \frac{1}{\tau^2} (g^{(0)}_{ab}+\tau^{2}g^{(2)}_{ab}+\tau^{3}g^{(3)}_{ab}+O(\tau^{4})) dx^a dx^b,
\end{equation}
where $\tau < 0$ denotes time and $x^{a}$ are the other coordinates. In terms of the conformal completion, $\tau=0$ is the future boundary $\scri^+$ of asymptotically de Sitter spacetimes and $x^{a}$ are the coordinates on $\scri^+$.
 Latin letters $a,b,\dots$ will always refer to three-dimensional indices of fields defined on $\scri^+$. The Starobinsky  expansion is related to the Fefferman-Graham expansion of asymptotically anti-de Sitter spacetimes by an analytical continuation. We call $g_{ab}^{(0)}$ the three-dimensional metric on the future boundary of asymptotically de Sitter $\scri^+$, $g^{ab}_{(0)}$ its inverse and we will denote as $D^{(0)}_a$ its metric-compatible covariant derivative. The holographic stress-energy tensor defined as $T_{ab}=\frac{3H}{16 \pi G}g_{ab}^{(3)}$ is trace-free and divergence-free with respect to the boundary metric $g^{(0)}_{ab}$ outside of sources: 
\begin{equation}\label{eq:Tab props}
    D^{(0)}_a T^{ab}=0 ,\qquad \ g_{(0)}^{ab} T_{ab}=0 \ ,
\end{equation}
where $T^{ab}$ is defined as $T_{ab}$ with indices raised with the inverse metric $g_{(0)}^{ab}$. 

In the presence of gravitational radiation, the boundary metric $g^{(0)}_{ab}$ generically admits no symmetry. In perturbation theory where $g_{\mu\nu}= \bar g_{\mu\nu}+\delta g_{\mu\nu}$ (where $\mu,\nu$ denote 4-dimensional indices), we can however write 
\begin{equation}
   g_{ab}^{(0)} = \bar  g_{ab}^{(0)} + \delta  g_{ab}^{(0)} ,
\end{equation}
where the background boundary metric $\bar  g_{ab}^{(0)}$, which describes the future boundary of de Sitter $\scri^+_\text{dS}$, admits symmetries. 
Motivated by the Bondi framework and by the description of spatially compact binary systems that reach future infinity $\scri^+$ at a single point $u=+\infty$, we choose $\bar  g_{ab}^{(0)}$ to be the metric on $\mathbb R \times S^2$
\begin{equation}
\bar g^{(0)}_{ab} dx^{a} dx^{b}= H^2 du^{2} + \mathring q_{AB}(x^C)dx^{A} dx^{B}. 
\end{equation}
Here $\mathring q_{AB}$ is the unit metric on $S^2$ and $u$ ranges over the real line. These coordinates do not cover the entire $\scri_\text{dS}^+$ of de Sitter, which has topology $S^3$, but $\scri_\text{dS}^+$ minus two points. The topology of $\scri^+$ depends upon the number of massive bodies. Here, we consider at least two such points removed at $u = \pm \infty$.  We also gauge fix the perturbation to so-called $\Lambda$-BMS gauge \cite{Compere:2019bua} such that 
\begin{equation}
g^{(0)}_{ab} dx^{a} dx^{b}= H^2 du^{2} + q_{AB}(u,x^C)dx^{A} dx^{B},
\end{equation}
and $\sqrt{q} = \sqrt{\mathring q}$. The inverse of $q_{AB}$ is denoted $q^{AB}$.  The shear is defined as $C_{AB}= H^{-2} \partial_u q_{AB}$. 

The background boundary metric admits 10 conformal Killing vectors $\bar \xi^a(u,x^A)$ which are in one-to-one correspondence with the Killing symmetries of the de Sitter background and define the $SO(1,4)$ group. By definition they obey 
\begin{align}\label{CKE}
   \bar g_{(0)c \langle a} \bar D_{(0) b \rangle} \bar \xi^c = 0. 
\end{align}
where $\bar D_{(0)a}$ is the covariant derivative compatible with the background boundary metric $\bar g_{(0)ab}$ and $\langle \cdot \rangle$ denote the symmetric tracefree part. The 10 conformal Killing vectors consist of the dilatations and rotations (which are Killing vectors of the background boundary metric) and the spatial translations and cosmological boosts (which are proper conformal Killing vectors of the background boundary metric). They  are given by the following table: 
\begin{center}
\begin{tabular}{ |c|c|c| } 
 \hline
 Charge & name  & generator \\ 
 \hline
 E & Dilatation & $\bar\xi^{u}=1, ~~\bar\xi^{A}=0$\\ 
 $P^{i}$ & Spatial translation & $\bar\xi^{u}_{(i)}=n_{i}\exp{(Hu)}, 
 ~~\bar\xi^{A}_{(i)}=-H\mathring{D}^{A}n_
 {i}\exp{(Hu)}$ \\ 
 $L^{i}$ & Rotation & $\bar\xi^{u}_{(i)}=0, 
 ~~\bar\xi^{A}_{(i)}=-\epsilon^{AB}\mathring{D}_{B}n_{i}$ \\
 $K^i$ & Cosmological boost & $\bar\xi^{u}_{(i)}=n_{i}\exp{(-Hu)}, 
 ~~\bar\xi^{A}_{(i)}=H\mathring{D}^{A}n_{i} \exp{(-Hu)}$ \\
 \hline
\end{tabular}
\end{center}
Here $n_i$ is the unit normal to the round sphere embedded in $\mathbb R^3$, and $\mathring{D}_{A}$ is the covariant derivative compatible with the unit sphere metric. 

For an arbitrary vector $\xi^a$, one has the identity
\begin{equation}
    \int_{M} du d^2\Omega \sqrt{\mathring q} g_{(0)}^{ac}D_{(0)c}( T_{ab}  \xi^b) =  \int_{M} du d^2\Omega  \sqrt{\mathring q} T^{ab}  g_{(0)c \langle a}  D_{(0) b \rangle}  \xi^c \ .
\end{equation}
for any region $M$ of $\scri^+$ outside of sources thanks to Eq. \eqref{eq:Tab props}. In perturbation theory, $T_{ab}$ admits linear and quadratic contributions in the perturbation $\delta g_{\mu\nu}$, $T_{ab}= \delta T_{ab}+\delta^2 T_{ab} + O((\delta g)^3)$. Let us now set $\xi^a = \bar \xi^a$ a background conformal Killing vector. Then, for any choice of spheres $S_1$ and $S_2$ such that $\partial M = S_1 \cup S_2$ (with opposite relative orientation), one has the equalities at linear and quadratic order, respectively,
\begin{align}\label{eq:abstract-conservation1}
    &\int_{S_1} d^2\Omega \sqrt{\mathring q} n^a \delta T_{ab} \bar \xi^b -  \int_{S_2} d^2\Omega \sqrt{\mathring q} n^a \delta T_{ab}  \bar \xi^b   =0,\\ 
\label{eq:abstract-conservation}
    &\int_{S_1} d^2\Omega \sqrt{\mathring q} n^a \delta^2 T_{ab} \bar \xi^b -  \int_{S_2} d^2\Omega \sqrt{\mathring q} n^a \delta^2 T_{ab}  \bar \xi^b   = \int_{M} du d^2\Omega \sqrt{\mathring q} \delta T^{ab} \mathcal L_{\bar \xi}  \delta g_{ab}^{(0)} ,
\end{align}
where $n^a$ is the unit normal to $S_1$ and $S_2$ defined as $n^a\partial_a = H^{-1}\partial_u$. The right-hand side of Eq. \eqref{eq:abstract-conservation1} is zero thanks to the conformal Killing equation \eqref{CKE}. 

Finally, we derived that on a given section of $\scri^+$ the charge defined from the holographic stress-energy tensor as
 \begin{eqnarray} \label{Charge}
    Q^T_{\bar \xi}(u) \equiv - \int_{S^2} d^2\Omega \sqrt{\mathring q}n^a T_{ab} \bar \xi^b
 \end{eqnarray}
  is a constant in linear theory, i.e. $Q_{\bar\xi}(u_1)=Q_{\bar\xi}(u_2)$ for any $u_1$, $u_2$. At quadratic order, the presence of radiation will induce a change in $Q_{\bar\xi}(u)$ given by the flux-balance  formula
\begin{align}
  &  Q^T_{\bar\xi} (u_2) -Q^T_{\bar\xi}(u_1) = \int_{u_1}^{u_2} du \, \dot Q^T_{\bar\xi} (u), \\
    \dot Q^T_{\bar \xi} (u) &\equiv \int_S d^2\Omega \sqrt{\mathring q}\delta T_{ab} \bar{g}^{ac}_{(0)} \bar{g}^{bd}_{(0)} \mathcal L_{\bar\xi} \delta g_{cd}^{(0)} + O((\delta g)^3).   
  \end{align}
  
Instead of considering the background symmetry generator $\bar \xi^a$ we could consider an arbitrary perturbation of the vector $\xi^a = \bar \xi^a + \delta \xi^a$. Assuming that the  $\Lambda$-BMS gauge is obeyed in the full theory, the generic infinitesimal diffeomorphism of the (4-dimensional) metric is parameterized by a vector $\xi^a$ of the boundary metric which obeys the equations
\begin{subequations}\label{eq5} 
 \begin{align}
\partial_u \xi^u &= \frac{1}{2 \sqrt{\mathring q} }\partial_A (\sqrt{\mathring q}  \xi^A ),\\ \label{eq6}
\partial_u \xi^A &= -H^2 q^{AB}\partial_B \xi^u . 
\end{align}   
\end{subequations}
Let us denote as $\xi^\mu$ the 4-dimensional vector associated with $\xi^a$ in $\Lambda$-BMS gauge. A linearized diffeomorphism of the 4-dimensional metric is simply the Lie derivative of the background metric with respect to $ \xi^\mu$, i.e. linearized gravitational fields are defined up to a gauge transformation $\delta g_{\mu\nu} \to \delta g_{\mu\nu} + \mathcal{L}_{\xi}\bar{g}_{\mu \nu}$.  We have $\mathcal L_{\xi} \bar g_{\mu\nu}=\mathcal L_{\delta \xi} \bar g_{\mu\nu}$ where the equality follows from the Killing equation $\mathcal L_{\bar \xi} \bar g_{\mu\nu} = 0$. Therefore, at linear order in perturbation theory and in $\Lambda$-BMS gauge, a perturbed vector $\delta \xi^a$ obeys the linearization of Eq. \eqref{eq5} where $q_{AB}$ is replaced by the background $\mathring q_{AB}$: 
\begin{subequations}\label{eq5bis} 
 \begin{align}
&\partial_u \delta \xi^u = \frac{1}{2 \sqrt{\mathring q} }\partial_A (\sqrt{\mathring q}  \delta \xi^A ) ,\\ 
&\partial_u \delta \xi^A =-H^2 \mathring q^{AB}\partial_B \delta \xi^u .
\end{align}   
\end{subequations}

For dilatations and rotations, we have $\partial_A \bar \xi^u = 0$ and we can fix $\xi^a=\bar\xi^a$ in the non-linear theory because the right-hand side of Eq. \eqref{eq6} is zero. We can therefore define universal (i.e. field independent) time translation and rotations that form a $\mathbb R \times SO(3)$ subalgebra of the $\Lambda$-BMS algebra in the non-linear theory. This subalgebra is part of the larger $\mathbb R \times \text{ADiff}(S^2)$ universal subalgebra of the $\Lambda$-BMS algebra which consists of time translations and area-preserving diffeomorphism (with generators $\xi^u=0$, $\xi^A = \varepsilon^{AB}\partial_B \Psi(\theta,\phi)$). The universal $\mathbb R \times SO(3)$ subalgebra of asymptotic symmetries was also identified in \cite{Bonga:2023eml} in a different boundary gauge. The background spatial translations and cosmological boosts are generically deformed in the non-linear theory. Yet, in the linear theory, the deformation vector is a background $\Lambda-$BMS generator (that obeys by definition \eqref{eq5bis}), which we can conventionally fix to be vanishing. With this convention, all background symmetries are undeformed in the linear theory.

We define the cosmological Bondi mass aspect $M^{(\Lambda)}$, angular momentum aspect $N_A^{(\Lambda)}$ and higher Bondi aspect $J_{AB}^{(\Lambda)}$ from the decomposition of the stress-energy tensor (see Eq. (3.15) of \cite{Compere:2019bua}) 
\begin{align}
   T_{ab}= \frac{3H}{16\pi G} \left[ \begin{array}{cc} -\frac{4}{3}M^{(\Lambda)} & -\frac{2}{3} N^{(\Lambda)}_B \\ -\frac{2}{3} N^{(\Lambda)}_B & J^{(\Lambda)}_{AB} + \frac{2}{\Lambda} M^{(\Lambda )} q_{AB} \end{array} \right]. 
\end{align}
Here $J^{(\Lambda)}_{AB}$ is traceless: $q^{AB}J^{(\Lambda)}_{AB} = 0$. The Bondi mass aspect $M$ and Bondi angular momentum aspect $N_A$ defined in the flat limit $H \mapsto 0$ are related to the cosmological quantities as (see Eqs. (2.38) and (2.52) of \cite{Compere:2019bua})
\begin{subequations}\label{defML}
   \begin{align}
   M^{(\Lambda )} & = M+\frac{1}{16} \partial_u (C_{CD} C^{CD}),\\ 
   N_A^{(\Lambda )} &= N_A - \frac{1}{2H^2} D^B N_{AB} + \frac{9}{32}\partial_A (C_{CD}C^{CD}). 
\end{align} 
\end{subequations}
The charges (for the vector $\xi^a$) can  be rewritten as 
\begin{eqnarray} \label{LBMS Charge}
    Q^T_\xi = \frac{1}{4\pi G} \int_{S^2} d^2\Omega \sqrt{\mathring q} \left[ M^{(\Lambda)}\xi^u + \frac{1}{2}N_A^{(\Lambda)}\xi^A \right] .
 \end{eqnarray}
In terms of Bondi aspects, the conservation of the holographic stress-energy tensor \eqref{eq:Tab props} is equivalent to 
\begin{subequations}\label{dM}
    \begin{align}
 \partial_u M^{(\Lambda)} +\frac{H^2}{2}D^A N_A^{(\Lambda)} +\frac{3H^4}{8}C_{AB}J^{{(\Lambda)}AB} &=0, \\
 \partial_u N_A^{(\Lambda)}-\partial_A M^{(\Lambda )}-\frac{3H^2}{2}D^B J_{AB}^{(\Lambda)} &=0. 
\end{align}
\end{subequations}
Here $D_{A}$ is the covariant derivative compatible with $q_{AB}$. Using Eqs. \eqref{dM} and \eqref{eq5}, the charge flux is then equivalently written as 
\begin{align} \label{30VIII24.01}
    \dot Q^T_\xi & = -\frac{3H^2}{32\pi G}\int d^2\Omega \sqrt{q}\left( H^2 \xi^u C_{AB} +2D_A \xi_B \right)J^{(\Lambda)AB}, \\
    &= -\frac{3H^2}{32\pi G}\int d^2\Omega \sqrt{q}\left( \xi^u \partial_u q_{AB} +\mathcal L_{ \mathring\xi}q_{AB} \right)J^{(\Lambda)AB},
\end{align}
where the Lie derivative is here understood to act on the $S^2$ manifold using the pullback vector $\mathring\xi = \xi^A \partial_A$.

The flux is quadratic in the metric perturbation. Upon expanding $\xi^a = \bar \xi^a +\delta \xi^a+O(\delta^2)$ (as well as its 2-dimensional pullback $\mathring \xi^A = \bar \xi^A +\delta \xi^A+O(\delta^2)$), $q_{AB}=\mathring  q_{AB}+\delta q_{AB}+O(\delta^2)$, $J^{(\Lambda)}_{AB}=\delta J^{(\Lambda)}_{AB}+\delta^2 J^{(\Lambda)}_{AB}+O(\delta^3)$ we have 
\begin{align}
    \delta J^{(\Lambda)}_{AB}  \mathring{q}^{AB} &=0 , \\
     \delta^2 J^{(\Lambda)}_{AB}  \mathring{q}^{AB} &=\delta J^{(\Lambda)}_{AB} \mathring q^{AC} \mathring q^{BD} \delta{q}_{CD},
\end{align}
as a consequence of the tracelessness of $J^{{(\Lambda)}AB}$, as well as 
\begin{align}
2\delta (D^{A}\bar \xi^{B}) \delta J_{AB}^{(\Lambda)}+2\delta^{2}J^{{(\Lambda)}AB}\mathring{D}_{B}\bar\xi_{A}&= (-\mathcal{L}_{\mathring{\bar\xi}}\delta{q}^{AB}+\mathring{D}_C \bar\xi^C \mathring q^{AC}\mathring q^{BD} \delta q_{CD}) \delta J_{AB}^{(\Lambda)}\nonumber \\
&= (\mathcal{L}_{\mathring{\bar\xi}}\delta{q}_{AB}-\mathring{D}_C  \bar\xi^C \delta q_{AB}) \mathring q^{AC} \mathring q^{BD} \delta J_{CD}^{(\Lambda)}.
\end{align}
The flux-formula therefore reads at quadratic level as 
\begin{align}\label{quadflux}
  \dot Q^T_{\xi} &= -\frac{3H^2}{32\pi G}\int d^2\Omega \sqrt{\mathring{q}} \mathring q^{AC} \mathring q^{BD}\left( \bar \xi^u \partial_u \delta q_{AB} +\mathcal L_{\mathring{\bar \xi}} \delta q_{AB}+\mathcal L_{\mathring{\delta \xi}}\mathring q_{AB}- \mathring{D}_E \bar \xi^E \delta q_{AB} \right)\delta J^{(\Lambda)}_{CD} +O(\delta^3)\ .
\end{align}
Here $\delta \xi^a$ obeys to Eqs. \eqref{eq5bis}. We choose the solution $\delta \xi^a=0$, i.e. we do not act with any further $\Lambda-$BMS symmetries but only with the background symmetry. In that sense, there are flux-balance laws associated with $SO(1,4)$ symmetry at quadratic order in perturbation theory.

In the flat limit, using the conventions of \cite{Compere:2023qoa}, the generators associated with time translations, spatial translations and rotations are simply the limit $H \mapsto 0$ of  $E$, $P^i$ and $L^i$, respectively. Lorenz boosts are associated with the limit $H \mapsto 0$ of the generator $ \frac{1}{2H}(K^i-P^i)$ with components $\bar \xi^{u}_{(i)}=-\frac{1}
 {H}n_{i}\sinh{(Hu)}$, $\bar \xi^{A}_{(i)}=\mathring{D}^{A}n_{i} \cosh{(Hu)}$. 
Using these relationships we notice that the charges $Q^T_\xi$ do not have the standard flat limit, see Eq. (3.2) of \cite{Barnich:2011mi}. In order to recover the standard flat limit, we will define instead the charges as 
\begin{align}
    Q_\xi(u) = Q_\xi^T (u) +\Delta Q_\xi (u) ,\label{shift}
\end{align}
where the shifting term $\Delta Q_\xi = O((\delta g)^2)$ will be defined in the next section. There is an ambiguity in defining this shift. We will partially fix this ambiguity by requiring that the energy is positive definite.

\section{The $SO(1,4)$ flux-balance laws}
\label{sec:eval}

In this section we will analyse the flux formulae at quadratic order in the quadrupolar approximation. We will use the quadrupolar solution derived in \cite{Compere:2023ktn}. In Bondi coordinates and $\Lambda$-BMS gauge, the metric reads as 
\begin{align}
ds^2 &= \left(\frac{\Lambda r ^{2}}{3}-\bigg(\frac{1}{2} R[q]+\frac{\Lambda}{12} C_{AB} C^{AB}\bigg)+\frac{2M}{r} +O(r^{-2})\right) du^{2} \nonumber \\
&+2\bigg(-1+\frac{1}{r^{2}}\frac{C_{AB}C^{AB}}{16}+\frac{1}{r^{4}}\frac{3}{16}\bigg(C^{AB} E_{AB}-\frac{3}{32} (C_{AB}C^{AB})^{2}\bigg)+O(r^{-6})\bigg)du dr \nonumber \\
&+2\bigg(\frac{1}{2} D^{B}C_{BA}+\frac{1}{r} \big(\frac{2}{3} N_{A}+\frac{1}{6} C_{A}^{F} D^{E}C_{EF}\big)+O(r^{-2})\bigg) dx^{A}du  \nonumber \\
&+ \bigg(r^2 q_{AB}+ r C_{AB}+\frac{1}{4} q_{AB }C_{CD}C^{CD}+\frac{E_{AB}}{r}+O(r^{-2})\bigg)dx^A dx^B . 
\end{align}
At linear order, we can expand the boundary metric $q_{AB}$, the shear $C_{AB}$ and the higher Bondi moment $E_{AB}$ as 
\begin{align}
q_{AB} = \mathring{q}_{AB}+ \delta q_{AB},\qquad 
C_{AB} = \delta C_{AB},\qquad 
E_{AB} = \delta E_{AB}. 
\end{align}

For convenience we introduce the notations 
\begin{align}
 \delta q_{AB}\equiv  e^i_{\langle A} e^j_{B \rangle} \delta q_{ij},\qquad \delta C_{AB}\equiv e^i_{\langle A} e^j_{B \rangle} \delta C_{ij} , \qquad \delta E_{AB}\equiv  e^i_{\langle A} e^j_{B \rangle} \delta E_{ij}.  
\end{align}
Here $r$ is the radial coordinate and $\theta^{A}=(\theta, \phi)$ are the angular coordinates. The unit normal vector is denoted as $n^{i}=x^{i}/r$. We employ the natural basis on the unit 2-sphere $e_{A}=\frac{\partial}{\partial \theta^{A}}$ embedded in $\mathbb{R}^{3}$ with components $e^{i}_{A}=\partial n^{i}/\partial \theta^{A}$. Given the unit sphere metric 
$\mathring{q}_{AB}=\mbox{diag}(1,\sin^{2} \theta)$, we 
have $n^{i}e^{i}_{A}=0$, $\mathring{q}_{AB}=\delta_{ij} 
e^{i}_{A}e^{j}_{B}$, and $\mathring{q}^{AB} 
e^{i}_{A}e^{j}_{B}=\perp^{ij}$, where 
$\perp^{ij}=\delta^{ij}-n^{i}n^{j}$ is the projector 
operator. Using the covariant derivative 
$\mathring{D}_{A}$ compatible with the unit sphere metric, 
$\mathring{D}_{A} \mathring{q}_{BC}=0$, we have 
$\mathring{D}_{A}e^{i}_{B}=\mathring{D}_{B}e^{i}_{A}=\mathring{D}_{A}\mathring{D}_{B}n^{i}=-\mathring{q}_{AB} 
n^{i}$. The transverse-traceless projector is 
denoted 
$\perp^{ijkl}_{\text{\footnotesize{tt}}}=\perp^{k(i}\perp^{j)l}-
\frac{1}{2}\perp^{ij} \perp^{kl}$. The tt projected part of any symmetric tensor $T_{ij}$ will be defined as $T^{\text{\footnotesize{tt}}}_{ij}=\perp^{ijkl}_{\text{\footnotesize{tt}}} T_{kl}$. We also use the notation $e^{i}_{\langle A}e^{j}_{B\rangle}=e^{i}_{(A}e^{j}_{B)}-\frac{1}{2}\mathring{q}_{AB}\perp^{ij}$ for the tracefree product of the basis vectors.

At quadrupolar order in the perturbation we can express the Bondi fields in terms of the even and odd source quadrupolar moments $Q_{ij}^{(\rho +p)}$ and $J_{ij}$ as \cite{Compere:2023ktn}
\begin{align}
G^{-1}\delta q_{ij} &=  \partial_u \zeta_{ij}+2H^2 \partial_u Q^{(\rho+p)}_{ij}+2H^2 n_k\epsilon_{kl(i}(K_{j)l}+H\int^u du'K_{j)l}(u')), \label{qABBondi}\\
G^{-1}\delta C_{ij} &=  3\zeta_{ij} +2 (\partial_u^2 -H^2)Q^{(\rho+p)}_{ij}+2n_k\epsilon_{kl(i} (\partial_u+H)K_{j)l},\label{CABBondi} \\
G^{-1}\delta E_{ij} & = 2  Q^{(\rho+p)}_{ij} + 2n_k\epsilon_{kl(i} J_{j)l}.  \label{EABBondi}
\end{align}
Here $K_{ij}=-(\partial_u - H)J_{ij}$ and 
$\zeta_{ij}$ obeys the differential equation 
\begin{align} \label{19XI24.02}
   \partial_u^2 \zeta_{ij}-3H^2 \zeta_{ij}=-2GH^{4}Q_{ij}^{(\rho + p)}.  
\end{align}
We further have 
\begin{align}\label{deltaM}
  G^{-1}  \delta M  &= Q^{(\rho)}-H P_{i \vert i}-3 n_i (P_i-H Q_i^{(\rho)}-H^2 P_{i \vert kk})+(3 n_i n_j - \delta_{ij})(\partial_u^2 Q_{ij}^{(\rho + p)}-H^2 Q_{ij}^{(\rho + p )}),\\
  G^{-1}  \delta N_i &= Q_i^{(\rho)}+H P_{i \vert kk}+n^j (\epsilon_{ijk}J_k +2 \partial_u Q_{ij}^{(\rho + p)})-2 \epsilon_{ijk} n_j n_l (K_{kl}- H J_{kl}),
\end{align}
where the linear Bondi mass aspect has been simplified as Eq. (5.35) of \cite{Harsh:2024kcl}.

\subsection{Energy loss}

From Eq. \eqref{quadflux}, the flux of the charge associated with dilatation reads at quadratic order in the perturbations as 
\bea \label{flux1}
\dot Q^T_{D} = -\frac{3H^{4}}{8G} \oint_S  \delta C_{MN} \delta 
J_{AB}^{(\Lambda)}~ \mathring{q}^{AM} \mathring{q}^{BN}.
\eea
Here we defined $\oint_S = \frac{1}{4\pi }\int_{S^2}d^2\Omega \sqrt{\mathring q}$.  The expression for $J^{(\Lambda)}_{AB}$ reads to linear order as \cite{Compere:2019bua}
\bea \nonumber
    3H^4\delta J^{(\Lambda)}_{AB}= - \partial_{u}  \delta N_{AB} -  H^2\bigg(\mathring D_{(A} \mathring D^{C}  \delta C_{B)C}-\frac{1}{2} \mathring {q}_{AB} \mathring{D}^{C} \mathring{D}^{D}  \delta C_{CD} -\delta C_{AB}\bigg) - 3H^4 \delta {E}_{AB} . \\ \label{12XII22.03}
\eea
Using the substitution \eqref{defML}, we can rewrite the flux-balance law of the dilatation \eqref{flux1} as 
\bea \label{14XI22.02} 
\partial_{u}\oint_S \delta^2 M &=-\frac{1}{8}\oint_S  \bigg[ \delta N_{AB}  \delta N_{CD} +H^{2} \delta C_{AB}( \delta C_{CD}-\mathring D_{C}\mathring D^{E}  \delta C_{DE})\nonumber\\
&-3H^{4}  \delta C_{AB}  \delta {E}_{CD}\bigg] \mathring{q}^{AC} \mathring{q}^{BD}. 
\eea
This formula also matches with the quadratic part of Eq. (2.52) of \cite{Compere:2019bua} after integrating over 2-sphere. Here we used that the linear energy $\oint_S \delta M$ is separately conserved, $\partial_u \oint_S \delta M=0$. From Eq. \eqref{deltaM}, the terms proportional to $n_i$ and $3n_i n_j-\delta_{ij}$ integrate to zero over the sphere while the term independent of $n_i$ is conserved as a consequence of the conservation of the stress-energy tensor, $\partial_u Q^{(\rho)}=-H Q^{(p)}=-H S_{ii}=H \partial_u P_{i \vert i}$ (see also Eq. (5.37) of \cite{Harsh:2024kcl}). Therefore, the energy loss only involves the quadratic perturbation $\delta^2 M$.

\subsubsection{Even parity energy loss}

We now ready to evaluate the flux-balance law of dilatations for quadratic modes. Let us first consider the even parity modes. The shear $C_{ij}$ is then independent of the angles. We can derive 
\bea 
\label{27VI24.01}
C_{AB}C^{AB} &=& \perp^{ijkl}_{\text{\footnotesize{tt}}} C_{ij} C_{kl}=C_{ij}^{\text{\footnotesize{tt}}}C_{ij} , \\ 
\label{27VI24.02}
\mathring{D}_{A}\mathring{D}^{C} C_{BC}&=& -
2(e^{i}_{(A}e^{j}_{{B)}}-n^{i}n^{j} 
\mathring{q}_{AB}) C_{ij},\\ \label{27VI24.03}
C^{AB}\mathring{D}_{A}\mathring{D}^{C} C_{BC}&=& - 
2C_{ij}^{\text{\footnotesize{tt}}} C_{ij}.
\eea
Equation \eqref{12XII22.03} simplifies to 
\bea \label{DeltaJAB}
\delta J^{(\Lambda)}_{AB}= -\frac{1}{3H^{4}} \partial_{u} \delta N_{AB} +\frac{1}{H^{2}} \delta C_{AB}-\delta E_{AB}.
\eea

Using the identities \eqref{27VI24.01}, \eqref{27VI24.02}, \eqref{27VI24.03}, the energy loss formula \eqref{14XI22.02} becomes
\bea \label{27VI24.04}
\partial_{u}\oint_S \delta^2 M =-\frac{1}{8G}\oint_S (\partial_{u} \delta C_{ij}^{\text{\footnotesize{tt}}} \partial_{u} \delta C_{ij}+ 3H^{2} \delta C_{ij}^{\text{\footnotesize{tt}}} \delta C_{ij} -3H^{4} \delta C_{ij}^{\text{\footnotesize{tt}}} \delta E_{ij}). 
\eea


Using the even-parity moments from \eqref{CABBondi}, \eqref{EABBondi}, the integrand of \eqref{27VI24.04} can be expanded as  
\begin{align} \nonumber
& G^{-2} \left(\partial_{u} \delta C_{ij}^{\text{\footnotesize{tt}}} \partial_{u} \delta C_{ij}+ 3H^{2} \delta C_{ij}^{\text{\footnotesize{tt}}} \delta C_{ij} -3H^{4} \delta C_{ij}^{\text{\footnotesize{tt}}} \delta E_{ij}\right) =\partial_{u}\bigg(9\zeta_{ij}^{\text{\footnotesize{tt}}} \dot\zeta_{ij}+12\ddot Q_{ij}^{(\rho+p)}\zeta_{{ij}}^{\text{\footnotesize{tt}}}\\ \nonumber
&-12H^{2} Q_{ij}^{(\rho+p)}\dot \zeta_{ij}^{\text{\footnotesize{tt}}}
-8H^{2}\ddot Q_{ij}^{(\rho+p)}\dot Q_{ij}^{(\rho+p)\text{\footnotesize{tt}}}-12H^{4} Q_{ij}^{(\rho+p)}\dot Q_{ij}^{(\rho+p)\text{\footnotesize{tt}}}\bigg) \\ \nonumber
&+4\bigg(\dddot Q_{ij}^{(\rho+p)}\dddot Q_{ij}^{(\rho+p)\text{\footnotesize{tt}}}+5H^{2}\ddot Q_{ij}^{(\rho+p)}\ddot Q_{ij}^{(\rho+p)\text{\footnotesize{tt}}}+4H^{4} \dot Q_{ij}^{(\rho+p)}\dot Q_{ij}^{(\rho+p)\text{\footnotesize{tt}}}\bigg).
\end{align}

We can finally rewrite the flux-balance law of dilatations as 
\begin{align} \label{23X24.01}
\partial_u Q^\text{e}_D(u) = -\frac{G}{2} \oint_S \bigg(\dddot Q_{ij}^{(\rho+p)}\dddot Q_{ij}^{(\rho+p)\text{\footnotesize{tt}}}+5H^{2}\ddot Q_{ij}^{(\rho+p)}\ddot Q_{ij}^{(\rho+p)\text{\footnotesize{tt}}}+4H^{4} \dot Q_{ij}^{(\rho+p)}\dot Q_{ij}^{(\rho+p)\text{\footnotesize{tt}}}\bigg),
\end{align}
where the dilatation charge in the even sector is identified as 
\begin{align}
Q^\text{e}_D &=G\oint_S \bigg( \frac{1}{G}\delta M+\frac{1}{G}\delta^2 M +\frac{9}{8}\zeta_{ij}^{\text{\footnotesize{tt}}} \dot\zeta_{ij}+\frac{3}{2} \ddot Q_{ij}^{(\rho+p)}\zeta_{{ij}}^{\text{\footnotesize{tt}}}
-\frac{3}{2}H^{2}Q_{ij}^{(\rho+p)}\dot \zeta_{ij}^{\text{\footnotesize{tt}}}
-H^{2}\ddot Q_{ij}^{(\rho+p)}\dot Q_{ij}^{(\rho+p)\text{\footnotesize{tt}}}\nonumber\\ 
&-\frac{3}{2}H^{4}Q_{ij}^{(\rho+p)}\dot Q_{ij}^{(\rho+p)\text{\footnotesize{tt}}} \bigg)   .  
\end{align}
The flux of energy is negative definite as it should. 
Using the identity 
\bea 
\oint \perp^{ijkl}_{\text{\footnotesize{tt}}}=\frac{2}{15} (3\delta^{k(i}\delta^{j)l}-\delta^{ij}\delta^{kl}),
\eea
the flux-balance law in \eqref{23X24.01} becomes
\bea \nonumber
\partial_u Q^\text{e}_D(u) &=& -\frac{G}{5}  \bigg((\dddot Q_{ij}^{(\rho+p)}-\frac{1}{3}\delta_{ij}\dddot Q^{(\rho+p)})^{2}+5H^{2}(\ddot Q_{ij}^{(\rho+p)}-\frac{1}{3}\delta_{ij}\ddot Q^{(\rho+p)})^{2} 
\nonumber \\ 
&+&4H^{4} (\dot Q_{ij}^{(\rho+p)}-\frac{1}{3}\delta_{ij}\dot Q^{(\rho+p)})^{2}\bigg)  .\label{evendM}
\eea

\subsubsection{Odd parity energy loss}

Let us now consider the odd parity modes. Denoting $C_{AB}=e^{i}_{\langle A}e^{j}_{B\rangle} C_{ij}$ with $C_{ij}=\epsilon_{ikl} n_{k} T_{jl}$, we now have
\bea 
\label{27VI24.01bis}
C_{AB}C^{AB} &=& \perp^{ijkl}_{\text{\footnotesize{tt}}} C_{ij} C_{kl}=C_{ij}^{\text{\footnotesize{tt}}}C_{ij} , \\ 
\label{27VI24.02bis}
\mathring{D}_{A}\mathring{D}^{C} C_{BC} &=& (4n^{(i}e^{j)}_{[A}e^{k}_{B]}-2n_{k} e^{i}_{(A}e^{j}_{B)})\epsilon_{ikl} T_{jl},\\ \label{27VI24.03bis}
\mathring{D}_{( A}\mathring{D}^{C} C_{B ) C} &=& -2C_{AB}, \\ 
C^{AB}\mathring{D}_{A}\mathring{D}^{C} C_{BC}&=& - 
2C_{ij}^{\text{\footnotesize{tt}}} C_{ij}.
\eea

Eq. \eqref{12XII22.03} again simplifies to Eq. \eqref{DeltaJAB} and  the energy loss formula \eqref{14XI22.02} becomes again \eqref{27VI24.04}.  Using the odd-parity moments from \eqref{CABBondi}, \eqref{EABBondi}, the integrand of \eqref{27VI24.04} becomes, 
\bea \nonumber
G^{-2}\left( \partial_{u} \delta C_{ij}^{\text{\footnotesize{tt}}} \partial_{u} \delta C_{ij}+ 3H^{2} \delta C_{ij}^{\text{\footnotesize{tt}}} \delta C_{ij} -3H^{4} \delta C_{ij}^{\text{\footnotesize{tt}}} \delta E_{ij}\right) =-8H^{2} \partial_{u}(\ddot J_{ij}^{\text{\footnotesize{tt}}} \dot J_{ij})-12H^{4}\partial_{u}(J_{ij}^{\text{\footnotesize{tt}}} \dot J_{ij}) \\ \nonumber
+4\bigg(\dddot J_{ij}^{\text{\footnotesize{tt}}}\dddot J_{ij}+5H^{2} \ddot J_{ij}^{\text{\footnotesize{tt}}}\ddot J_{ij}+4H^{4} \dot J_{ij}^{\text{\footnotesize{tt}}}\dot J_{ij}\bigg).
\eea
We can finally rewrite the flux-balance law of dilatations in the odd sector as 
\begin{align}
\partial_u Q^\text{o}_D(u) = -\frac{G}{2} \oint_S \bigg( \dddot J_{ij}^{\text{\footnotesize{tt}}}\dddot J_{ij}+5H^{2} \ddot J_{ij}^{\text{\footnotesize{tt}}}\ddot J_{ij}+4H^{4} \dot J_{ij}^{\text{\footnotesize{tt}}}\dot J_{ij} \bigg),\label{odddM}
\end{align}
where the dilatation charge is identified as 
\begin{align}
Q^\text{o}_D &=\oint_S \bigg( \frac{1}{G}\delta^2 M -G H^2 \ddot J_{ij}^{\text{\footnotesize{tt}}}\dot J_{ij}- \frac{3 G H^4}{2}  J_{ij}^{\text{\footnotesize{tt}}}\dot J_{ij} \bigg)   .  
\end{align}
Finally, the total dilation charge $Q_D$ is the sum of the even and odd sectors: 
\begin{align}
Q_D &=G\oint_S \bigg( \frac{1}{G}\delta M+\frac{1}{G}\delta^2 M +\frac{9}{8}\zeta_{ij}^{\text{\footnotesize{tt}}} \dot\zeta_{ij}+\frac{3}{2} \ddot Q_{ij}^{(\rho+p)}\zeta_{{ij}}^{\text{\footnotesize{tt}}}
-\frac{3}{2}H^{2}Q_{ij}^{(\rho+p)}\dot \zeta_{ij}^{\text{\footnotesize{tt}}}
-H^{2}\ddot Q_{ij}^{(\rho+p)}\dot Q_{ij}^{(\rho+p)\text{\footnotesize{tt}}}\nonumber\\ 
&-\frac{3}{2}H^{4}Q_{ij}^{(\rho+p)}\dot Q_{ij}^{(\rho+p)\text{\footnotesize{tt}}} -G H^2 \ddot J_{ij}^{\text{\footnotesize{tt}}}\dot J_{ij}- \frac{3 G H^4}{2}  J_{ij}^{\text{\footnotesize{tt}}}\dot J_{ij}\bigg)   .  
\end{align}
It obeys the dilatation flux balance law 
\begin{align}\label{dotQD}
 \dot Q_D &= \partial_u Q^\text{e}_D + \partial_u Q^{\text{o}}_D,    
\end{align}
where the even and odd energy fluxes are given in Eqs. \eqref{23X24.01}, \eqref{odddM}. 

\subsection{Angular momentum loss}
From \eqref{quadflux}, the quadratic part of the angular momentum loss formula becomes,
\bea 
\dot Q^T_{L^i}=-\frac{3H^{2}}{8G} \oint_S  \mathring{q}^{AM} \mathring{q}^{BN}  
 {\mathcal{L}}_{\mathring{\bar \xi}_{(i)}}\delta q_{MN}\, \delta J^{(\Lambda)}_{AB} . 
\eea

Using Eq. \eqref{DeltaJAB}, the angular momentum loss formula can be written as
\bea \nonumber
&&\partial_{u}\bigg(Q^T_{L^i}-\frac{1}{8 G H^{2}} \oint_S     
\left( {\mathcal{L}}_{\mathring{\bar\xi}_{(i)}}\delta q_{MN} \right) \partial_{u}\delta C_{AB} 
~\mathring{q}^{AM} \mathring{q}^{BN}\bigg)=-\frac{1}{8G} \oint_S  \bigg[\left( {\mathcal{L}}_{\mathring{\bar\xi}_{(i)}}\delta C_{MN}\right) \partial_{u} \delta C_{AB}\\ 
&& +3 \left( {\mathcal{L}}_{\mathring{\bar\xi}_{(i)}}\delta q_{MN}\right) \delta C_{AB}
-3H^{2} \left( {\mathcal{L}}_{\mathring{\bar\xi}_{(i)}}\delta q_{MN}\right) \delta E_{AB}
\bigg]~\mathring{q}^{AM} \mathring{q}^{BN}.  \label{Jflux}
\eea

Let us first discuss the even parity modes. We note the identities $\epsilon^{AB}=e^A_i e^B_j n_k \epsilon_{ijk}$, $\oint_S 3 n_i n_j=\delta_{ij}$ and $\oint_S n_i n_j n_k n_l = \frac{1}{15}(\delta_{ij}\delta_{kl}+\delta_{ik}\delta_{jl}+\delta_{il}\delta_{jk})$. After performing the integral over the sphere, the right-hand side of Eq. \eqref{Jflux} becomes 
\bea 
&&-\frac{1}{10G} \epsilon_{i}{}^{mn} \bigg( \delta C_{mk} \partial_{u} \delta C_{kn}+3\delta q_{mk}  \delta C_{kn}-3H^{2} \delta q_{mk}  \delta E_{kn}\bigg).
\eea
After some algebra we can rewrite the latter expression as 
\bea 
&-\frac{G}{10}\epsilon_{i}{}^{mn}\partial_{u}\bigg(6\zeta_{km} \ddot Q_{kn}^{(\rho+p)}-6H^{2} \zeta_{km} Q_{kn}^{(\rho+p)}
+6 \dot \zeta_{kn}\dot Q_{km}^{(\rho+p)}-4H^{2} Q_{km} \ddot Q_{kn}^{(\rho+p)}\bigg)\nonumber  \\ 
&-\frac{2G}{5}\epsilon_{i}{}^{mn}\bigg(\ddot Q_{km}^{(\rho+p)}\dddot Q_{kn}^{(\rho+p)}+5H^{2} \dot Q_{km}^{(\rho+p)}\ddot Q_{kn}^{(\rho+p)}+4H^{4}  Q_{km}^{(\rho+p)}\dot Q_{kn}^{(\rho+p)}\bigg). 
\eea

Let us now consider the odd parity modes. After performing the integral over the sphere and reorganizing the terms, the right-hand side of Eq. \eqref{Jflux} becomes  
\bea 
&-\frac{G}{10}\epsilon_{i}{}^{mn}\bigg[\partial_{u}\bigg(-4H^{2}J_{km} \ddot J_{kn}-12H^{4}\int^{u} J_{mk} \dot J_{kn}\bigg) \nonumber \\ 
&+4\bigg(\ddot J_{km}\dddot J_{kn}+5H^{2} \dot J_{km}\ddot J_{kn}+4H^{4}  J_{km}\dot J_{kn}\bigg)\bigg].
\eea
Combining both even and odd parity sectors, the flux-balance law of angular momentum finally reads as 
\begin{align} \label{2XI24.01}
\partial_u Q_{L^i} = -\frac{2G}{5} \epsilon_{imn}  \bigg( & \ddot Q_{km}^{(\rho+p)}\dddot Q_{kn}^{(\rho+p)}+5H^{2} \dot Q_{km}^{(\rho+p)}\ddot Q_{kn}^{(\rho+p)}+4H^{4}  Q_{km}^{(\rho+p)}\dot Q_{kn}^{(\rho+p)} \nonumber \\ 
 & + \ddot J_{km}\dddot J_{kn}+5H^{2} \dot J_{km}\ddot J_{kn}+4H^{4}  J_{km}\dot J_{kn}\bigg),
\end{align}
where the final  angular momentum charge (in both parity sectors) is 
\begin{align}
Q_{L_i} &=Q^T_{L_i}-\frac{1}{8 G H^{2}} \oint_S     
\left( {\mathcal{L}}_{\mathring{\bar\xi}_{(i)}}\delta q_{MN} \right) \partial_{u}\delta C_{AB} 
~\mathring{q}^{AM} \mathring{q}^{BN} \nonumber \\  \nonumber
&+\frac{G}{10 }\epsilon_{imn}   \bigg( 6\zeta_{km} \ddot Q_{kn}^{(\rho+p)}-6H^{2} \zeta_{km} Q_{kn}^{(\rho+p)}
+6 \dot \zeta_{kn}\dot Q_{km}^{(\rho+p)}-4H^{2} Q_{km} \ddot Q_{kn}^{(\rho+p)}\\ 
&-4H^{2}J_{km} \ddot J_{kn}-12H^{4} \int^{u} J_{km} \dot J_{kn}\bigg). 
\end{align}

\subsection{Linear momentum and boost charge losses}
From Eq. \eqref{quadflux} with $\delta \xi^a =0$, the quadratic part of the linear momentum loss and cosmological boost formula becomes,
\bea 
\dot Q^T_{\bar \xi_{(i)}}=-\frac{3H^{2}}{8 G} \oint_S  \bigg(H^{2}\bar\xi^{u}_{(i)} \delta C_{MN}+ \mathcal{L}_{\mathring{\bar\xi}}\delta q_{MN} - \mathring D_C \bar\xi_{(i)}^C \delta q_{MN}\bigg) \delta J^{(\Lambda)}_{AB} {}\mathring{q}^{AM} \mathring{q}^{BN},
\eea
where $\bar\xi^{u}_{(i)}=n_{i}\exp(\varepsilon Hu)$; $\bar\xi^{A}_{(i)}=-\varepsilon H \exp(\varepsilon Hu)\mathring{D}_{A}n_{i}$. For linear momenta $\varepsilon =1$ while for cosmological boosts $\varepsilon=-1$. We simplify the expression as 
\bea \label{5XI24.03}
\dot Q^T_{\bar \xi_{(i)}}=-\frac{3H^{2}}{8G} \oint_S \exp(\varepsilon Hu)\bigg(H^{2}n_{i} \delta C_{MN}-\varepsilon H (\mathring{D}^{C}n_{i})(\mathring{D}_{C}\delta q_{MN})\bigg) \delta J_{AB}^{(\Lambda)} {}\mathring{q}^{AM} \mathring{q}^{BN}.
\eea
In order to perform the angular integrals, we note that for any symmetric tracefree tensors $V_{AB}$, $W_{AB}$ that can be decomposed into even and odd parts as 
\begin{align}
   V_{AB} =e_{\langle A}^i e_{B \rangle}^j (V_{ij}^e + \epsilon_{ikl}n_k V_{lj}^o ), \\
   W_{AB} =e_{\langle A}^i e_{B \rangle}^j (W_{ij}^e + \epsilon_{ikl}n_k W_{lj}^o),  
\end{align}
we have the integrals 
\begin{align}
    & \oint_S n^i V_{AB} W^{AB} = -\frac{4}{15} \epsilon_{imt} \left( V_{ml}^e W_{tl}^o + W_{ml}^e V_{tl}^o \right) ,\\
    & \oint_S e^C_i \mathring{D}_C V_{AB} W^{AB}= -\frac{4}{15} \epsilon_{imt} \left( V_{ml}^e W_{tl}^o + W_{ml}^e V_{tl}^o \right) . 
\end{align}
Therefore, 
\begin{align}
    \dot{Q}^T_{\bar \xi_{(i)}} &= \frac{H^2}{10G} \exp(\varepsilon Hu) \epsilon_{imt}
    \left[ (H^2 \delta C_{ml}^e-\varepsilon H \delta q_{ml}^e) \delta J^{(\Lambda)o}_{tl} +  (H^2 \delta C_{tl}^o-\varepsilon H \delta q_{tl}^o)\delta J^{(\Lambda)e}_{ml} \right].
\end{align}
We now substitute $\delta J^{(\Lambda)}_{ij}$ using Eq. \eqref{DeltaJAB} and integrate by parts to obtain 
\begin{align} \label{15XI24.01}
\dot Q_{\bar \xi_{(i)}}^{\text{int}} &= \frac{\exp(\varepsilon Hu)}{10G} \epsilon_{imt} \bigg[   \frac{1}{3}\bigg(2\delta N_{ml}^{e} \delta N_{tl}^{o}+6H^{2} \delta C_{ml}^e \delta C_{tl}^o-3H^{4} \delta C_{ml}^e \delta E_{tl}^{o}-3H^{4} \delta E_{ml}^e \delta C_{tl}^{o}\bigg) \nonumber\\ 
&-\frac{\varepsilon H}{3}  \delta q_{ml}^e\bigg( 3\delta C_{lt}^o-3H^{2}\delta E_{tl}^o\bigg) -\frac{\varepsilon H}{3}  \delta q_{lt}^o\bigg( 3\delta C_{ml}^e-3H^{2}\delta E_{ml}^e\bigg)\nonumber \\
    & - \frac{1 }{3} \bigg(\delta q_{ml}^e \delta N_{tl}^o+\delta N_{ml}^e \delta q_{lt}^o\bigg)   \bigg],
\end{align}
where the charge (defined as an intermediate quantity for now, hence the superscript) is 
\bea 
\hspace{-.5cm}Q_{\bar \xi_{(i)}}^{\text{int}}  = Q_{\bar \xi_{(i)}}^T +\frac{1}{30G}  \exp(\varepsilon Hu) \epsilon_{imt}  \left[\delta C_{ml}^e \delta N_{tl}^o + \delta N_{ml}^e \delta C_{tl}^o -\frac{\varepsilon}{H}(\delta q_{ml}^e \delta N_{lt}^o+\delta N_{ml}^e \delta q_{lt}^o)\right].
\eea
Substituting all Bondi fields in terms of multipolar moments, we rewrite the flux-balance law as 
\begin{align} \label{BoostFlux}
\dot Q_{\bar \xi_{(i)}} = -\frac{4G e^{\varepsilon Hu}}{{15}} \epsilon_{imt} (3 \varepsilon H^{3}  \zeta_{ml} \dot J_{lt}+2H^{6} Q_{ml} J_{lt}+ 6H^{4} \dot Q_{ml} \dot J_{lt}+6H^{2} \ddot Q_{ml} \ddot J_{lt}+ \dddot Q_{ml} \dddot J_{lt}),
\end{align}
where the final charge is defined as 
\begin{align} \nonumber
Q_{\bar \xi_{(i)}} &= Q_{\bar \xi_{(i)}}^{\text{int}} +\frac{e^{\varepsilon Hu}}{15} \epsilon_{imt}\bigg[6H^{2}\partial_{u}(\dot Q_{ml} \dot J_{lt}) + 8 H^{4}\partial_{u}(Q_{ml} J_{lt})-4 \varepsilon H^{5} \partial_u \left( Q_{ml} \int^u J_{lt} \right)\\ \label{FinalBoostCharge}
&\hspace{-1cm}
-5 \dot \zeta_{ml} \ddot J_{lt}+ 3H^{2} (\dot \zeta_{ml} J_{lt}-\zeta_{ml} \dot J_{lt})+8\varepsilon H \dot \zeta_{ml} \dot J_{lt} -\bigg(2H^{4}\ddot Q_{ml}-6H^{6} Q_{ml}-3 \varepsilon H^{3}\dot \zeta_{ml}
\bigg)\int^u J_{lt}
\bigg] . 
\end{align}

\subsection{Flat spacetime limit}

Let us now prove that we recover the standard Poincar\'e flux-balance laws in the quadrupolar approximation upon taking the flat spacetime limit. In the quadrupolar truncation, these laws read as \cite{Compere:2019gft}\footnote{Note that $\mathcal G_i$ in \cite{Compere:2019gft} is $\mathcal N_i$ in our conventions, see Section 9 of \cite{Compere:2023qoa}.} (original derivations can be found in   \cite{1975ApJ...197..717E,PhysRev.128.2471,1973ApJ...183..657B,RevModPhys.52.299,Kozameh:2017qiw})  
\begin{align}
\partial_u \mathcal E &= - \frac{G}{5}\dddot I_{ij}^{\rm PM} \dddot I_{ij}^{\rm PM}  -  \frac{16G}{45} \dddot J_{ij}^{\rm PM}  \dddot J_{ij}^{\rm PM}    ,\\
\partial_u \mathcal P_i &=  -  \frac{16G}{45} \epsilon_{ijk} \dddot I_{jl}^{\rm PM} \dddot J_{kl}^{\rm PM}     ,\\
\partial_u \mathcal J_i &=  - \frac{2G}{5} \epsilon_{ijk} \ddot I_{jl}^{\rm PM} \dddot I_{kl}^{\rm PM}   - \frac{32G}{45} \epsilon_{ijk} \ddot J_{jl}^{\rm PM} \dddot J_{kl}^{\rm PM}       ,\\
\partial_u \mathcal N_i  &= \mathcal P_i    .\label{dotNi}
\end{align}
where the flat tensors $I_{ij}^{\rm PM}$,  $I_{ij}^{\rm PM}$ are symmetric and tracefree. 
The charges are expressed as the flat limit of the $SO(1,4)$ generators as $\mathcal E = \mathcal Q_{D}$, $\mathcal P_i =Q_{P^i}$, $\mathcal J_i = Q_{L^i}$ and $\mathcal K_i =\text{lim}_{H \mapsto 0} \frac{1}{2H}( Q_{K^i} - Q_{P^i}) $ with $\mathcal N_i = \mathcal K_i + u \mathcal P_i$. 

The flat limit of Eqs. \eqref{evendM}-\eqref{odddM}-\eqref{2XI24.01} reproduce the flat flux-balance laws of energy and angular momentum with the dictionary 
\begin{align}
    I^{\rm PM}_{ij} &= Q_{ij}^{(\rho +p)} - \frac{1}{3}\delta_{ij} Q_{kk}^{(\rho + p)},\\
    J^{\rm PM}_{ij} &= \frac{3}{4}J_{ij}. 
\end{align}

The flat limit of Eq. \eqref{BoostFlux} reproduces the flat flux-balance law for the momentum. For Lorentz boosts we have to extract the $O(H)$ term in Eq. \eqref{15XI24.01}. Note that $\delta q_{ab} = O(H^{2})$, therefore, only the  $\exp(\varepsilon Hu) \delta N_{ml}^{e} \delta N_{tl}^{o}$ term in Eq. \eqref{15XI24.01} will contribute to the boost charge. From Eq. \eqref{BoostFlux} we obtain 
\bea 
&&\dot {\mathcal K}_i =\text{lim}_{H \mapsto 0} \frac{1}{2H}( \dot Q_{K^i} - \dot Q_{P^i})= \frac{4Gu}{15} \epsilon_{iml} \dddot Q_{mj} \dddot J_{jl}.
\eea
We can finally shift the charge as 
\begin{align}
 \mathcal N_i = \mathcal K_i    -\frac{4Gu}{15} \int^u \epsilon_{iml} (\dddot Q_{mj} \dddot J_{jl} )
\end{align}
to obtain the correct flux-balance law \eqref{dotNi}.


\section{Comparison with the literature}
\label{sec:comp}

The first proposed quadrupolar formula for the energy flux for even parity modes was derived in \cite{Ashtekar:2015lxa}, see also \cite{Date:2016uzr}. In our earlier work \cite{Compere:2023ktn}, we pointed out that there is an inconsistent quadrupolar truncation in \cite{Ashtekar:2015lxa, Date:2016uzr}, which leads to incorrect terms proportional to $Q_{ij}^{(p)}$ even in the even parity sector. Performing the quadrupolar truncation using the methods of \cite{Compere:2023ktn}, let us now upgrade the flux formula of \cite{Ashtekar:2015lxa} to include the $Q_{ij}^{(p)}$ terms. We first note the algebraic identity 
\bea \nonumber
\dddot Q_{{ij}}^{(\rho+p)} \dddot Q_{{ij}}^{\text{\footnotesize{tt}}(\rho+p)}+5H^{2} \ddot Q_{ij}^{(\rho+p)} \ddot Q_{ij}^{\text{\footnotesize{tt}}(\rho+p)}+4H^{4} \dot Q_{ij}^{(\rho+p)} \dot Q_{ij}^{\text{\footnotesize{tt}}(\rho+p)} = \mathcal{R}^{\text{e}}_{ij} \mathcal{R}^{\text{e \footnotesize{tt}}}_{ij} -\partial_u \delta E^{\text{e}} ,
\eea
where we defined 
\bea 
\mathcal{R}^{\text{e}}_{ij} &=& \dddot Q_{ij}^{(\rho+p)}-3H \ddot Q_{ij}^{(\rho+p)}+ 2H^{2} \dot Q_{ij}^{(\rho+p)}, \\
\delta E^{\text{e}} &=& -3H \ddot Q_{ij}^{(\rho+p)} \ddot Q_{ij}^{\text{\footnotesize{tt}}(\rho+p)} +4H^{2} \dot Q_{ij}^{(\rho+p)} \ddot Q_{ij}^{\text{\footnotesize{tt}}(\rho+p)}
-6H^{3} \dot Q_{ij}^{(\rho+p)} \dot Q_{ij}^{\text{\footnotesize{tt}}(\rho+p)}.
\eea
The same identity holds in the odd sector with $Q_{ij}^{(\rho +p)}$ replaced by $J_{ij}$, namely,
\bea \nonumber
\dddot J_{ij} \dddot J_{ij}^{\text{\footnotesize{tt}}}+5H^{2} \ddot J_{ij} \ddot J_{ij}^{\text{\footnotesize{tt}}}+4H^{4} \dot J_{ij} \dot J_{ij}^{\text{\footnotesize{tt}}}= \mathcal{R}^{\text{o}}_{ij} \mathcal{R}^{\text{o \footnotesize{tt}}}_{ij} -\partial_u \delta E^{\text{o}} ,
\eea
where we defined 
\bea 
\mathcal{R}^{\text{o}}_{ij} &=& \dddot J_{ij} -3H \ddot J_{ij} + 2H^{2} \dot J_{ij} , \\
\delta E^{\text{o}} &=& -3H \ddot J_{ij} \ddot J_{ij}^{\text{\footnotesize{tt}}} +4H^{2} \dot J_{ij}  \ddot J_{ij}^{\text{\footnotesize{tt}}}
-6H^{3} \dot J_{ij}  \dot J_{ij}^{\text{\footnotesize{tt}}}.
\eea

The energy flux balance law \eqref{dotQD}  can therefore be equivalently defined as 
\begin{align}
\partial_u \left( Q_D +\frac{G}{2}\oint_S \delta (E^{\text{e}} +E^{\text{o}})\right) &=  -   \frac{G}{2} \oint_S  \left( \mathcal{R}^{\text{e}}_{ij}\mathcal{R}_{ij}^{\text{e \footnotesize{tt}}} + \mathcal{R}^{\text{o}}_{ij}\mathcal{R}_{ij}^{\text{o \footnotesize{tt}}}\right) \nonumber \\
& = -  \frac{G}{5}   \left( \mathcal{R}^{\text{e}}_{ij}\mathcal{R}^{\text{e}}_{ij}+\mathcal{R}^{\text{o}}_{ij}\mathcal{R}^{\text{o}}_{ij}\right).\label{energyflux2}
\end{align}
The right-hand side is negative definite. It depends upon the combination $Q_{ij}^{(\rho+p)}$, consistently with the Bondi fields \eqref{qABBondi}, \eqref{CABBondi}, \eqref{EABBondi}, \eqref{deltaM}. The energy flux \eqref{energyflux2} is our upgrade of the formula proposed in \cite{Ashtekar:2015lxa}\footnote{Note that the power radiated quadrupolar formula in \cite{Ashtekar:2015lxa} was given in terms of the transverse-traceless (TT) part of $\mathcal{R}_{ij}$ field. Let us recall that any symmetric rank-2 tensor can be decomposed as
$Q_{ij}=\frac{1}{3} \delta_{ij} \delta^{kl} Q_{kl}+(\partial_{i}\partial_{j}-\frac{1}{{3}} \delta_{ij} \nabla^{2})B+\partial_{i}B_{j}^{T}+\partial_{j}B_{i}^{T}+Q_{ij}^{TT}$ where $Q_{ij}^{TT}$ is the TT part of $Q_{ij}$, which satisfies $\partial^{i} Q_{ij}^{TT}=0= \delta^{ij} Q_{ij}^{TT}$. The two notions, $\text{\footnotesize{tt}}$ projection and TT part of a rank-2 symmetric tensor are therefore generically distinct. However, this distinction disappears after integrating over the two sphere at infinity \cite{Dobkowski-Rylko:2024jmh, Hoque:2017xop}. Note in addition that in \cite{Ashtekar:2015lxa}, $\dot Q_{ij}$ denotes the Lie derivative with respect to dilatation Killing vector field (T) in conformal coordinates of de Sitter, i.e. $\dot Q_{ij}=\mathcal{L}_{T}Q_{ij}$. After conversion to Bondi frame, it becomes $\mathcal{L}_{T}Q_{ij}=(\partial_{u}-2H)Q_{ij}$.}.

We can also upgrade the angular momentum flux proposed in \cite{Ashtekar:2015lxa}. Again we note the identity 
\bea 
\ddot Q_{km}^{(\rho+p)} \dddot Q_{kn}^{(\rho+p)}+5H^{2}\dot Q_{km}^{(\rho+p)} \ddot Q_{kn}^{(\rho+p)}+ 4H^{4}  Q_{km}^{(\rho+p)} \dot Q_{kn}^{(\rho+p)}= \mathcal{L}_{km}^{\text{e}}\mathcal{R}_{kn}^{\text{e}}-\partial_{u}\delta L^{\text{e}},
\eea
where we defined 
\bea 
\mathcal{L}_{km}^{\text{e}}&=& \ddot Q_{km}^{(\rho+p)}-3H \dot Q_{km}^{(\rho+p)}+2H^{2} Q_{km}^{(\rho+p)},\\ \nonumber
\delta L^{\text{e}} &=& -3H \dot Q_{km}^{(\rho+p)} \ddot Q_{kn}^{(\rho+p)}+2H^{2} Q_{km}^{(\rho+p)} \ddot Q_{kn}^{(\rho+p)}+ 2H^{2} \dot Q_{km}^{(\rho+p)}\dot Q_{kn}^{(\rho+p)} -6H^{3} Q_{km}^{(\rho+p)} \dot Q_{kn}^{(\rho+p)}.\\
\eea
The same identity holds in the odd sector with $Q_{ij}^{(\rho+p)}$ replaced by $J_{ij}$,
\bea 
\ddot J_{km} \dddot J_{kn}+5H^{2}\dot J_{km} \ddot J_{kn}+ 4H^{4}  J_{km} \dot J_{kn}= \mathcal{L}_{km}^{\text{o}}\mathcal{R}_{kn}^{\text{o}}-\partial_{u}\delta L_{T}^{\text{o}},
\eea
where we defined 
\bea 
\mathcal{L}_{km}^{\text{o}}&=& \ddot J_{km}-3H \dot J_{km}+2H^{2} J_{km},\\ 
\delta L^{\text{o}} &=& -3H \dot J_{km} \ddot J_{kn}+2H^{2} J_{km} \ddot J_{kn}+ 2H^{2} \dot J_{km}\dot J_{kn} -6H^{3} J_{km} \dot J_{kn}.
\eea
Hence the angular momentum flux balance law \eqref{2XI24.01} can also be equivalently defined as 
\bea 
\partial_{u}\bigg(Q_{L^{i}}-\frac{2G}{5}\epsilon_{imn}(\delta L^{\text{e}}+\delta L_{T}^{\text{o}})\bigg)=-\frac{2G}{5}\epsilon_{imn} (\mathcal{L}_{km}^{\text{e}} \mathcal{R}_{kn}^{\text{e}}+\mathcal{L}_{km}^{o} \mathcal{R}_{kn}^{\text{o}}).
\eea
This is our upgrade of the angular momentum flux proposed in \cite{Ashtekar:2015lxa}. 

In anti-de Sitter spacetime with Dirichlet boundary conditions, the $SO(3,2)$ conserved quantities are defined using the holographic stress-energy tensor \cite{Balasubramanian:1999re}. More precisely, they are defined from the analytic continuation to negative cosmological constant of the charge formula \eqref{Charge}. We started from this charge formula as initial ansatz in the case of positive cosmological constant in this paper. When rewritten in Bondi variables, the charge can be expressed in terms of the $\Lambda$-corrected Bondi mass aspects $M^{(\Lambda)}$ and Bondi angular momentum aspects $N_A^{(\Lambda)}$ while the flux is proportional to the field $J^{(\Lambda)}_{AB}$ \cite{Compere:2019bua,Compere:2020lrt}, see Eq. \eqref{30VIII24.01}. However, for leaky boundary conditions, where the boundary metric is allowed to vary, as it is the case here, the holographic stress-tensor does not allow by itself to define the quasi-conserved quantities. Indeed, computing the flux of $M^{(\Lambda)}$ \eqref{defML} using Eq. \eqref{27VI24.04} shows that the ``would be energy flux'' associated with the Bondi mass is not manifestly negative definite. It cannot therefore lead to a proposal for a definition of energy in the presence of a positive cosmological constant. 

 Bonga, Bunster, and Pérez (BBP) \cite{Bonga:2023eml} studied several aspects of gravitational waves in de Sitter for both linear and the non-linear theory. For the study of linearised solutions they focused on solving the homogeneous Einstein equations in a partial wave expansion to the gravitational field. One of the results of their paper is the derivation of an energy flux formula for $l=2$ mode of linearized perturbations around de Sitter. Starting from the energy loss formula in a different boundary gauge fixing introduced in \cite{Chrusciel:2020rlz, Chrusciel:2021ttc}, they obtained a quadratic energy flux formula for linearized pertubations in de Sitter background. 
 In \cite{Harsh:2024kcl}, an energy flux formula for de Sitter Teukolsky waves has also been obtained. With the correct identification of the variables, the energy flux formula for BBP quadrupolar waves exactly matches with that of de Sitter Teukolsky waves. In this section, we wish to compare our quadrupole formula with the energy flux formula for BBP quadrupolar waves in de Sitter \cite{Bonga:2023eml}.

For $l=2$ even parity linearized solutions of \cite{Bonga:2023eml}, the energy flux is given by
\bea 
E^{(E)} =-\frac{{3}}{8\pi G} \sum_{m=-2}^{m=+2}\int_{u=-\infty}^{u=+\infty}\bigg[|\dddot a_{m}|^{2}+5H^{2}|\ddot a_{m}|^{2}+4H^{4}|\dot a_{m}|^{2}\bigg] du,
\eea
where $a_{m}=a_{m}(u)$ is a function of retarded time $u$. The function $a_{m}(u)$ can also be understood as the coefficient of the $l=2$ even parity solution of homogeneous linearized Einstein's equation. The relationship between the variables $a_{m}$ used in \cite{Bonga:2023eml} and our even parity quadrupole moments are obtained from \cite{Harsh:2024kcl}, where a relation between the BBP wave solutions and $l=2$ mode quadrupolar truncated solutions \cite{Compere:2023ktn} have been achieved via de Sitter Teukolsky waves. Using Eqs. (3.55), (5.79) of 
\cite{Harsh:2024kcl}, we obtain
\bea \label{19XI24.01}
a_{m}=\frac{H^{2}}{2}q_{m}-\frac{1}{6} \ddot q_{m},
\eea
where $q_{m}$ can be written in terms of $l=2$
spherical harmonics $Y^{2m}(\theta,\phi)=\mathcal{Y}^{2m}_{ij}n_{i}n_{j}$,
\bea 
q_{m}=\frac{1}{{H^{4}}}  \frac{24\pi}{15} (\mathcal{Y}_{ij}^{2m})^{*} \zeta_{ij}(u).
\eea
An explicit expression for $\mathcal{Y}^{2m}_{ij}$ can be found in Eq. (3.218) of \cite{Maggiore:2007ulw}.
Therefore using \eqref{19XI24.01}, and the identity \eqref{19XI24.02}, we have
\bea
a_{m}=\frac{8\pi G}{{15}} (\mathcal{Y}_{ij}^{2m})^{*} Q_{ij}^{(\rho+p)}.
\eea

Therefore,
\bea 
\sum_{m=-2}^{m=+2} |\dddot a_{m}|^{2}&=&\frac{8\pi G^{2}}{15} \bigg(\dddot Q_{ij}^{(\rho+p)} \dddot Q_{ij}^{(\rho+p)}-\frac{1}{3}\dddot Q^{(\rho+p)}_{ij}\dddot Q^{(\rho+p)}_{kl}\delta_{ij}\delta_{kl}\bigg),\\
&=& \frac{8\pi G^{2}}{15} \bigg(\dddot Q_{ij}^{(\rho+p)}-\frac{1}{3}\dddot Q^{(\rho+p)}\delta_{ij}\bigg)^{2},
\eea
where we have used the identity,
\bea \label{28VI24.01}
\sum_{m=-2}^{m=+2} (\mathcal{Y}_{ij}^{2m})^{*} (\mathcal{Y}_{kl}^{2m})= \frac{15}{16\pi} (\delta_{ik}\delta_{jl}+\delta_{il}\delta_{jk}-\frac{2}{3}\delta_{ij}\delta_{kl}).
\eea
Hence the energy flux in terms of even-parity quadrupole moments becomes
\bea \nonumber
E^{(E)}=-\frac{G}{5} \int_{-\infty}^{+\infty} du \bigg((\dddot Q_{ij}^{(\rho+p)} -\frac{1}{3}\dddot Q^{(\rho+p)} \delta_{ij})^{2}+5H^{2} (\ddot Q_{ij}^{(\rho+p)} -\frac{1}{3}\ddot Q^{(\rho+p)} \delta_{ij})^{2} \\
+4H^{4}(\dot Q_{ij}^{(\rho+p)} -\frac{1}{3}\dot Q^{(\rho+p)} \delta_{ij})^{2}\bigg).
\eea
This can also be written as 
\bea 
\hspace{-.8cm}E^{(E)}=-\frac{G}{8\pi} \int_{-\infty}^{+\infty}du\int_{S^{2}} d\Omega \bigg(\dddot Q_{ij}^{(\rho+p)}\dddot Q_{ij}^{(\rho+p)\text{\footnotesize{tt}}} +5H^{2} \ddot Q_{ij}^{(\rho+p)} \ddot Q_{ij}^{(\rho+p)\text{\footnotesize{tt}}}
+4H^{4}\dot Q_{ij}^{(\rho+p)} \dot Q_{ij}^{(\rho+p)\text{\footnotesize{tt}}}\bigg)\!.
\eea
Considering now the $l=2$ odd parity linearized solution, the energy flux is given by 
\bea 
E^{(B)} =-\frac{{3}}{8\pi G} \sum_{m=-2}^{m=+2}\int_{u=-\infty}^{u=+\infty}\bigg[|\dddot b_{m}|^{2}+5H^{2}|\ddot b_{m}|^{2}+4H^{4}|\dot b_{m}|^{2}\bigg] du,
\eea
where $b_{m}=b_{m}(u)$ is a function of retarded time $u$. The relationship between the variable $b_{m}$  used in \cite{Bonga:2023eml} and our odd parity $l=2$ mode quadrupole moments are given by \cite{Harsh:2024kcl},
\bea 
b_{m}=-\frac{8\pi G}{{15}} (\mathcal{Y}_{ij}^{2m})^{*} J_{ij}.
\eea
Therefore using \eqref{28VI24.01}, we have
\bea 
\sum_{m=-2}^{m=+2} |\dddot b_{m}|^{2}&=&\frac{8\pi G^{2}}{15} \dddot J_{ij} \dddot J_{ij}.
\eea
Note that by construction $J_{ij}$ is symmetric and traceless.
Hence the energy flux in terms of odd-parity quadrupole moments becomes
\bea 
E^{(B)}=-\frac{G}{5} \int_{-\infty}^{+\infty} du \bigg(\dddot J_{ij}\dddot J_{ij}+5H^{2} \ddot J_{ij}\ddot J_{ij}+ 4H^{4} \dot J_{ij}\dot J_{ij}\bigg).
\eea
This can also be written as
\bea 
E^{(B)}=-\frac{G}{8\pi} \int_{-\infty}^{+\infty}du\int_{S^{2}} d\Omega\bigg(\dddot J_{ij}\dddot J_{ij}^\text{\footnotesize{tt}}+5H^{2} \ddot J_{ij}\ddot J_{ij}^\text{\footnotesize{tt}}+ 4H^{4} \dot J_{ij}\dot J_{ij}^\text{\footnotesize{tt}}\bigg).
\eea
We have therefore demonstrated that the energy flux formulae \eqref{23X24.01}, \eqref{odddM} exactly coincide with the ones derived in \cite{Bonga:2023eml}. This is remarkable because these formulae have been obtained in different boundary gauges. In \cite{Bonga:2023eml}, the angular components $g_{AB}$ of the boundary metric were kept fixed, while here we kept fixed the determinant of $g_{AB}$ and the mixed components $g_{uA}$. This does not prove that the energy flux formulae are gauge invariant, but they are at least invariant under an interesting class of gauge transformations.

Given that the quadratic energy loss formula derived in \cite{Chrusciel:2020rlz} and \cite{Kolanowski:2020wfg} agree with the one derived in \cite{Bonga:2023eml} as shown in \cite{Bonga:2023eml} we also proved the equivalence of the energy flux formulae \eqref{23X24.01}, \eqref{odddM} with \cite{Chrusciel:2020rlz,Kolanowski:2020wfg}.

\section{Conclusion}
\label{sec:ccl}

We derived the full set of flux-balance laws for linear (even and odd) quadrupolar perturbations associated with background $SO(1,4)$ symmetries. These laws reduce to the standard Poincar\'e flux-balance laws in the flat limit.  Even though the flux-balance laws have been determined, their split into the time derivative of the charge (left-hand side) and the flux (right-hand side) is not unique. We warn the reader that even if the symmetry generators form a $SO(1,4)$ algebra, we did not prove that the charges form a $SO(1,4)$ algebra under a suitable bracket. As proven in \cite{Compere:2020lrt}, the charges $Q^T_\xi$ defined from the holographic stress-energy tensor represent the $\Lambda$-BMS algebra under the adjusted bracket \cite{Barnich:2011mi}. However, since the charges are here defined with a shift \eqref{shift}, one would need to study the impact of this shift on the representation theorem. This  remains an open question. 

After analysis, we obtained two distinct proposals for the energy loss. Both proposals admit the standard flat limit and are negative definite at quadratic level. They differ by terms proportional to the Hubble radius. The first proposal is given in Eq. \eqref{energyflux2}. It corresponds to an upgrade of the formula first proposed in \cite{Ashtekar:2015lxa} after correcting the even parity sector following \cite{Compere:2023ktn}, and after including the odd parity contribution. The second proposal is given as Eq. \eqref{dotQD}. After integrating over $\scri^+$, it matches with the total energy loss formula derived in \cite{Bonga:2023eml,Harsh:2024kcl}. We further complemented the previous literature \cite{Bonga:2023eml,Harsh:2024kcl} by providing the corresponding definition of energy and energy flux on a fixed cut (i.e. fixed $u$) of $\scri^+$ in terms of multipolar moments. The existence of two proposals for the energy loss indicates that fundamental requirements are missing to uniquely single out one expression. One such requirement is gauge invariance. The Bondi framework is based upon one gauge choice and even though it was successful in the asymptotically flat regime, we expect that a gauge-independent framework is required to settle this issue. The comparison of proposed definitions of energy loss based on Weyl scalars and Killing spinors \cite{Szabados:2015wqa,Saw:2016isu,Saw:2017amv} with the current framework is left for future work.


\section*{Acknowledgements}

We gratefully thank A. Pérez and A. Virmani for their comments on the draft prior to submission. G.C. is Research Director of the F.R.S.-FNRS. The work of JH is supported in part by MSCA Fellowships CZ - UK2 $(\mbox{reg. n. CZ}.02.01.01/00/22\_010/0008115)$
from the Programme Johannes Amos Comenius co-funded by the European Union. JH also acknowledges the support from Czech Science Foundation Grant 22-14791S. ESK is supported by Scuola Normale Superiore and partially by INFN (IS-GSS) and ERC (NOTIMEFORCOSMO, 101126304) grant funded by the European Union. Views and opinions expressed are, however, those of the author(s) only and do not necessarily reflect those of the European Union or the European Research Council Executive Agency. Neither the European Union nor the granting authority can be held responsible for them. ESK has also been supported by TUBITAK-2219 Fellowship during the project.

\bibliography{dS_memory}
\end{document}